\documentstyle[fleqn,12pt]{article}
\begin{document}

\begin{titlepage}
\vspace*{1.9cm}

\begin{center}
{\LARGE   \bf $U(1)$ gauge theory over discrete space-time 
                 and phase transitions}
\end{center}
\vspace*{0.5cm}

\begin{center}
{\bf Liangzhong Hu} \\

\vspace*{0.5cm}
{\small{\it Institute of Mathematics, Peking University,}}
{\small{\it     Beijing 100871, P. R. China   }} \\
\end{center}
\vspace*{0.5cm}

\begin{abstract}
We first apply Connes' noncommutative geometry to a finite point space. The
explicit form of the action functional of $U(1)$ gauge field on this $n$-point
space is obtained. We then consider the case when the $n$-point space is 
replaced by $\{$space-time$\}$ $\times$ $\{n$-point space$\}$. This action 
is shown to relate the Hamiltonian of the continuous-spin formulation of the 
Potts model. We argue that $U(1)$ gauge theory on the discrete space-time 
determines the geometric origin of a class of phase transitions.
\end{abstract}
\vspace*{0.2cm}
\vspace*{0.1cm}
\end{titlepage}

{\bf 1.  Introduction}

Within the framework of Connes' noncommutative geometry \cite{Co1,Co2}, the Higgs field
and the symmetry breaking mechanism in the standard model have a remarkable
geometrical picture. The Higgs field is a connection, which arises from the
geometry of the two-point space \cite{Co3,Co4}. (See also 
\cite{Coq,Kas,Cham1,Var1,Din,Co5,Co6,Cham2,Song,Cham3,Var2}, and the references 
therein.) This suggests that the discrete geometry may play an important role 
in physics. Differrential calculus and gauge theory on discrete groups was 
proposed in \cite{Si1}. The systematic approach towards the differential 
calculus and the gauge theory on arbitrary  finite or countable sets was 
formulated in \cite{Di1,Di2}.
It is natural to ask: what is the explicit form of the action functional
of gauge fields on the $n$-point space? Here $n$ is the natural number ($n\geq 2$).
When $n=2$, the action is just the case of Higgs field.
The purpose of the present paper is to answer this question for the simplest
case, i.e., the gauge group is taken to be $U(1)$.

In Section 2 we will review differential calculus on $n$-point
space which is formulated by Dimakis and M\"{u}ller-Hoissen \cite{Di1,Di2},
and then in Section 3 we will construct the spectral triple on $n$-point
space and derive the explicit form of action functional of $U(1)$ gauge fields. 
In section 4 the action funcional over $\{$space-time$\}$ $\times$ $\{ n$-point  
space$\}$ is obtained. In the concluding section, we comment on the physical 
meaning of the action functional we derived and its relation to the theory of 
phase transitions.

{\bf 2. Differential calculus on $n$-point space}

In this section we shall review the differential calculus on $n$-point space.
More detailed account of the construction can be found in \cite{Di1,Di2}.

Let $M$ be a set of $n$ points $i_{1}, \cdots, i_{n}$ $(n<\infty)$, and
$\cal A$ the algebra of complex functions on $M$ with
$(fg)(i) = f(i)g(i)$.  Let $p_{i} \in \cal{A}$ defined by
\begin{equation}
p_{i}(j) = \delta_{ij}.            \label{2.1}
\end{equation}
Then it follows that
\begin{equation}
p_{i} p_{j} = \delta_{ij} p_{j},  \ \ \ \ \ \ \ \
\sum_{i} p_{i} = 1.                           \label{2.2}
\end{equation}
In other words, $p_{i}$ is a projective operator in $\cal A$.
Each $f \in \cal A$ can be written as
\begin{equation}
f = \sum_{i} f(i) p_{i},                                \label{2.3}
\end{equation}
where $f(i) \in \bf C$, a complex number.
The algebra $\cal A$ can be extended to a universal differential algebra
$\Omega(\cal A)$. The differentials satisfy the following relations:
\begin{eqnarray}
p_{i} d p_{j} &=& -(d p_{i}) p_{j}
                      + \delta_{i j} d p_{i},     \label{2.4}\\
\sum_{i} d p_{i} &=& 0.                                     \label{2.5}
\end{eqnarray}
This means that the differential calculus over $n$-point space $M$ associates
with it $n-1$ linear independent differentials.
There is a natural geometrical representation associated with $M$.
Let the projective operators $p_{i}$ $(i=1, \cdots, n)$ be the orthonormal
base vectors in the Euclidean space ${\bf R}^{n}$. Then $M$ forms the vertices
of the $(n-1)$-dimensional hypertetrahedron embeded in ${\bf R}^{n}$.

The universal first order differential calculus $\Omega^{1}$ is generated
by $p_{i} d p_{j} (i \ne j)$, $i, j = 1, 2, \cdots, n$.
Notice that $p_{i} d p_{i}$ is the linear combinations of
$p_{i} d p_{j} (i \ne j)$.

Similarly, the compositions of
 $p_{i} d p_{j} (i \ne j)$, $i, j = 1, 2, \cdots, n$ generate the higher
order universal differential calculus on $M$.
For example,
the universal second order differential calculus $\Omega^{2}$ is generated
by $p_{i} d p_{j} p_{j} d p_{k} (i \ne j, j \ne k)$,
$i, j, k = 1, 2, \cdots, n$.

A simple calculation shows that 
\begin{equation}
d p_{i} = \sum_{j}(p_{j} d p_{i} - p_{i} d p_{j}).            \label{2.01}
\end{equation}
Furthermore,
\begin{equation}
d p_{i} d p_{j}= \sum_{k}(p_{k} d p_{i} p_{i} d p_{j} 
                   - p_{i} d p_{k} p_{k} d p_{j} 
                   + p_{i} d p_{j} p_{j} d p_{k}).            \label{2.02}
\end{equation}
Any $1$-form $\alpha$ can be written as
$\alpha = \sum_{i,j} \alpha_{ij} p_{i} d p_{j}$
with $\alpha_{ij} \in \bf C$ and
$\alpha_{ii} =0$.  One then finds
\begin{equation}
d \alpha = \sum_{i,j,k}(\alpha_{jk} - \alpha_{ik} + \alpha_{ij})
 p_{i} d p_{j} p_{j} d p_{k}.                   \label{2.6}
\end{equation}
Now we consider a connection $\alpha$ on $M$, $\alpha$ is a $1$-form, and
skew -adjoint, $\alpha^{*} = -\alpha$.
$\alpha$ obeys the usual transformation rule,
\begin{equation}
\alpha' = u\alpha u^{*} +u d u^{*}.                           \label{2.7}
\end{equation}
Here $u = \sum_{i} u(i) p_{i} \in \cal {A}$, and $u(i) \in U(1)$, the Abelian
unitary group. $\alpha$ is thus called the $U(1)$ gauge field on $M$.
In order to make the formulae concise, we introduce
\begin{equation}
a = \sum_{i,j} (1 + \alpha_{ij}) p_{i} d p_{j}.             \label{2.8}
\end{equation}
Notice that
\begin{equation}
a_{ii} = 1.                         \label{2.9}
\end{equation}
One then has
\begin{eqnarray}
a' &=& u a u^{*}, \nonumber   \\
a_{ij}^{'} &=& u(i) a_{ij} u(j)^{*}.                           \label{2.10}
\end{eqnarray}
The curvature of the connection $\alpha$ is given by
\begin{equation}
\theta = d \alpha + \alpha^{2},                                         \label{2.11}
\end{equation}
and transforms in the usual way, $\theta' = u \theta u^{*}$.
As a $2$-form, $\theta$ can be written as
\begin{eqnarray}
\theta &=& \sum_{i,j,k} \theta_{ijk} p_{i} d p_{j} p_{j} d p_{k},  \nonumber \\
\theta_{ijk} &=& a_{ij} a_{jk} - a_{ik}.              \label{2.12}
\end{eqnarray}

{\bf 3. From spectral triple to action functional over $M$}

In noncommutative geometry, all the geometrical data is determined
by a spectral triple $({\cal {A}}, {\cal {H}}, D)$, where $\cal{A}$ is an
involutive algebra, $\cal {H}$ is a Hilbert space with an involutive
representation $\pi$ of $\cal A$, $D$ is a self-adjoint operator acting on
$\cal H$.

We now construct the spectral triple $({\cal {A}}, {\cal {H}}, D)$
over the $n$-point space $M$. In our case, $\cal A$ is the algebra on $M$
defined in the last section. Without loss of generality, $\cal H$ is taken
to be the $n$-dimensional linear space over $\bf C$, 
i.e., $\cal H$ is just the direct sum 
${\cal H}  = \bigoplus_{i=1}^{n} {\cal H}_{i}$, 
${\cal H}_{i} = \bf {C}$. The action of $\cal {A}$ is given by
\begin{equation}
 f\in {\cal A} \longrightarrow 
               \left ( \begin{array}{cccc}
                        f(1)      &    0 & ... & 0   \\
                        0      &    f(2) & ... & 0    \\
                         ...   &     ... &  ... & ... \\
                         0     &   0    & ...   & f(n)
                               \end{array}  \right ).
\end{equation}
Then $D$ is the Hermitian $n \times n$ matrix with elements 
$D_{ij} =D_{ji}^{*}$,
and $D_{ij}$ is a linear mapping from ${\cal H}_{j}$ to ${\cal H}_{i}$.
We have the following equality defines the involutive representation of
$ d a$ ($a \in \cal A)$ in $\cal H$,
\begin{equation}
 \pi(d a) = i[D, \pi(a)].                     \label{3.1}
\end{equation}
To ensure the differential $ d$ satisfies
\begin{equation}
       d^{2} =0,                                    \label{3.3}
\end{equation}
one has to impose the following condition on $D$,
\begin{equation}
D^{2} = {\mu}^{2} I,                                  \label{3.4}
\end{equation}
where $\mu$ is a real constant and $I$ is the $n \times n$ unit matrix.

We now take $D_{ij} \ne  0$ $(i\ne j).$ Then the representation 
$\pi:\; \Omega(\cal{A}) \rightarrow \cal{L}(\cal{H})$ is injective  on 
$\Omega^{1}(\cal{A})$. 

In Connes' language \cite{Co2}, Our spectral triple $({\cal A}, {\cal H}, D)$
is odd (except the case of $n = 2$). For the sake of convenience,
we omit the symbol $\pi$ from now on.

The projective operator $p_i$ can be expressed as the $n \times n$ matrix now,
\begin{equation}
(p_{i})_{\alpha \beta} = \delta_{\alpha i} \delta_{\beta i}.    \label{3.5}
\end{equation}
Notice that $D$ commute exactly with the action of $\cal A$, for the sake of
convenience, we can ignore the diagonal elements of $D$, i. e.,
\begin{equation}
D_{ii} = 0.                                         \label{3.2}
\end{equation}
From (\ref{3.1}) and (\ref{3.5}), one has
\begin{eqnarray}
(p_{i} d p_{j})_{\alpha \beta} &=& i \delta_{\alpha i} \delta_{\beta j}
                                         D_{ij},    \label{3.6}  \\
(p_{i} d p_{j} p_{j} d p_{k} p_{k} d p_{l} p_{l} d p_{r})_{\alpha \beta}
        &=& \delta_{\alpha i} \delta_{\beta r}
                D_{ij} D_{jk} D_{kl} D_{lr}.  \label{3.7}
\end{eqnarray}
One can define an inner product $< \vert >$  in $\Omega^{*} \cal A$ by setting
\begin{equation}
<\alpha\vert\beta> =tr_{\omega}(\alpha^{*}\beta\vert D \vert^{-p}). \label{3.100}
\end{equation}
where $tr_{\omega}$ is the Dixmier trace. In our case $ p = 0$ and 
$tr_{\omega}$ is reduced to the ordinary trace,
\begin{equation}
<\alpha\vert\beta> =tr(\alpha^{*}\beta).      \label{3.8}
\end{equation}
Since the gauge field $\alpha$ satisfies $\alpha^{*} =-\alpha$, 
one has $\theta^{*}=\theta$.
Then the action functional of $\theta$ is
\begin{equation}
S=\Vert \theta \Vert^{2} =<\theta \vert \theta>=tr \theta^{2}.                    \label{3.9}
\end{equation}
From (\ref{2.12}), (\ref{3.7}) and (\ref{3.9}), we have
\begin{equation}
S=\sum_{i,j,k,l} \theta_{ijk} \theta_{kli}D_{ij} D_{jk} D_{kl} D_{li}.  \label{3.10}
\end{equation}
Denote
\begin{equation}
a_{ij} D_{ij} = H_{ij},                    \label{3.11}
\end{equation}
where $a_{ij}$ is defined in (\ref{2.8}).
Then $H = (H_{ij})$ is a Hermitian matrix with
\begin{equation}
H_{ii} = 0.                                  \label{3.12}
\end{equation}
From (\ref{2.12}), (\ref{3.4}), (\ref{3.10}), (\ref{3.11}) and (\ref{3.12}), 
one thus has
\begin{equation}
S = trH^{4} -2{\mu}^{2} trH^{2} +n{\mu}^{4}.   \label{3.13}
\end{equation}
From (\ref{3.12}), the eigenvalues $\lambda_{i} \;\; (i=1,2,\cdots,n)$
of $H$ satisfy:
\begin{equation}
 \sum^{n}_{i=1} \lambda_{i} = 0.                                 \label{3.101}
\end{equation}
(\ref{3.13}) can be written as the following:
\begin{equation}
S = \sum^{n}_{i=1} \lambda_{i}^{4} 
    -2{\mu}^{2} \sum^{n}_{i=1} \lambda_{i}^{2} +n{\mu}^{4}.     \label{3.102}
\end{equation}
The eigenvalues of such kind of $H$'s generate $(n-1)$-dimensional Euclidean 
space ${\bf R}^{n-1}$. By the quadric transformation, (\ref{3.102}) can be
taken the form as
\begin{equation}
S = C_{2}^{ijkl}\sum^{n-1}_{i,j,k,l=1} \varphi_{i}\varphi_{j}
                                        \varphi_{k}\varphi_{l} 
    -C_{1} \sum^{n-1}_{i=1} \varphi_{i}^{2} +n{\mu}^{4},     \label{3.103}
\end{equation}
where $(\varphi_{1},\cdots,\varphi_{n-1})$ is a vector in ${\bf R}^{n-1}$, 
$C_{2}^{ijkl}$ and $C_{1}$ are real constants.

For the sake of convenience, we identify ${\bf R}^{n-1}$ with a subspace
embedded in the $n$-dimensional geometrical
representation space of $M$ introduced in Section 2 from now on. 
In the $(n-1)$-dimensional rectangular coordinate system, the reference point 
is taken to be the center of the $(n-1)$-dimensional hypertetrahedron. 
$M$ can then be represented by a set of $n$ vectors in ${\bf R}^{n-1}$: 
$e^{\alpha}_{i}$ ($\alpha = 1, \cdots, n$; $i=1, \cdots, n-1$), such that
\begin{equation}
\sum_{i} e^{\alpha}_{i} e^{\beta}_{i} =\frac{n}{n-1} \delta^{\alpha \beta}
             -\frac{1}{n-1}.                                        \label{3.15}
\end{equation}
In (\ref{3.15}) we have chosen the normalization of the vectors to be unity
for convenience. This set of $e$'s satisfy
\begin{eqnarray}
              \sum_{\alpha} e^{\alpha}_{i} &=& 0;    \label{3.16}      \\
\sum_{\alpha} e^{\alpha}_{i} e^{\alpha}_{j}
                    &=& \frac{n}{n-1} \delta_{ij}.    \label{3.17}
\end{eqnarray}
It should be mentioned that the properties of $M$ is encoded in those of the
set of spin states in the Potts model \cite{Zia}. The reason will be discussed 
in section 5.
Using (\ref{3.16}), the eigenvalues of $H$ is
\begin{equation}
\lambda_{\alpha} = \sum^{n-1}_{i=1} \phi_{i} e^{\alpha}_{i},
 \ \ \ \ \ \ \alpha=1, \cdots, n.          \label{3.18}
\end{equation}
Here $\phi_{i} (i=1, \cdots, n-1)$ is a real parameter.
We call ${\bf\Phi} = (\phi_{1}, \cdots, \phi_{n-1})$ the order 
parameter field in ${\bf R}^{n-1}$.
Finally from (\ref{3.13}), (\ref{3.17}) and (\ref{3.18}), we obtain the explicit
form of $S$ over the $n$-point space $M$:
\begin{equation}
S = \sum_{i,j,k,l} (\sum_{\alpha} e^{\alpha}_{i} e^{\alpha}_{j} e^{\alpha}_{k}
     e^{\alpha}_{l}) \phi_{i} \phi_{j} \phi_{k} \phi_{l}
    -\frac{2n}{n-1} {\mu}^{2} (\sum_{i} \phi^{2}_{i}) + n {\mu}^{4}.
                                                              \label{3.19}
\end{equation}
Notice that the constant term $n {\mu}^{4}$ can be left out.

When $n=2$, (\ref{3.19}) changes into
\begin{equation}
S =2(\phi^{2} -\mu^{2})^{2}.        \label{3.20}
\end{equation}
This is just the Hamiltonian density of the Landau phenomenological theory of
phase transitions below the critical temperature \cite{Land}.
Here $\phi$ is known as the order parameter. Notice that the size of
the coefficients of $\phi^{2}$ and $\phi^{4}$ does not affect the values of
critical exponents of phase transitions, but it may affect the mass
value of the Higgs field when (\ref{3.20}) is considered as the Higgs potential:
From (\ref{3.4}), (\ref{3.11}) and (\ref{3.18}), we have $\phi^{2} =
 \mu^{2} \vert a_{12} \vert^{2}$. One then has
\begin{equation}
S = 2\mu^{4}(\vert a_{12} \vert^{2} -1)^{2},       \label{3.21}
\end{equation}
which is the form of Connes' version of Higgs potential.

{\bf 4. Action functional over $V \times M$}

Now we construct the $U(1)$ gauge field theory over $\{$space-time$\}$ 
$\times$ $\{n$-point space$\}$. 
Denote space-time by $V$, 
$\cal A$ the algebra of complex functions on $V \times M$. 
Just as in Section 2, Each  $f \in \cal A$ can be written as
\begin{equation}
f = \sum_{i} f(i) p_{i}.                                \label{4.1}
\end{equation}
Notice that this time $f(i) $ is a complex function over $V_{i}$,
the $ith$ copy of $V$.
 $\cal A$ can be also extended to a universal differential algebra
$\Omega(\cal A)$.  Denote the differential on $M$ by $d_{f}$. In other words,
The differential $d$ in Section 2 and Section 3 is replaced by 
$d_{f}$. Let $d_{s}$ be the differential on $V$, and $d$ the total 
differential on $V \times M$. One then has
\begin{equation}
 d = d_{s} + d_{f}.                        \label{4.2}
\end{equation}
The nilpotency of $d$ requires that
\begin{equation}
 d_{s} d_{f} = - d_{f} d_{s}.                        \label{4.3}
\end{equation}
Differentiate (\ref{4.1}), we have
\begin{equation}
d f = \sum_{i} (d_{s} f(i)) p_{i} +\sum_{i} f(i) d_{f} p_{i}.   \label{4.4}
\end{equation}
Any $1$-form $\alpha$ can be written as
\begin{equation}
\alpha = \sum_{i,j} \alpha_{ij} p_{i} d_{f} p_{j} + 
                                      \sum_{i} \alpha_{i} p_{i},  \label{4.5}
\end{equation}
with $\alpha_{ij}$, the complex function on $V$ and
$\alpha_{ii} =0$; $\alpha_{i}$, the $1$-form on $V_{i}$.  One then finds
\begin{eqnarray}
d \alpha &=& \sum_{ij}(d_{s} \alpha_{ij})p_{i} d_{f} p_{j}  
 + \sum_{i,j,k}(\alpha_{jk} - \alpha_{ik} + \alpha_{ij})
 p_{i} d_{f} p_{j} p_{j} d_{f} p_{k}         \\ \nonumber
     & &\mbox{}   
        + \sum_{i}(d_{s}\alpha_{i})p_{i} 
        -\sum_{i}\alpha_{i} d_{f} p_{i} .                   \label{4.6}
\end{eqnarray}
Now we consider a connection $\alpha$ on $V \times M$, $\alpha$ is a $1$-form 
and skew-adjoint, i.e., $\alpha$ is given by (\ref{4.5}) and $\alpha^{*} 
= -\alpha$. $\alpha$ obeys the usual transformation rule,
\begin{equation}
\alpha' = u \alpha u^{*} +u d u^{*}.                           \label{4.7}
\end{equation}
Here $u = \sum_{i} u(i) p_{i} \in \cal {A}$, and $u(i) \in U(1)$, the Abelian
unitary group on $V_{i}$. $\alpha$ is thus called the $U(1)$ gauge field on 
$V \times M$. In this section, we only consider the simpliest case, i.e.,   
\begin{eqnarray}
\alpha_{i} &=& A,                            \label{4.8}    \\
    u(i)   &=& u(j)  \;\;\;\;\;\;\; (i,j =1,2, \cdots, n).    \label{4.9}
\end{eqnarray}
Here $A$ is a $U(1)$ gauge field on $V$.  
The physical meaning of the above asumptions is: there exist unique gauge 
field, i.e., the Maxwell electromagnetic field over all copies of $V$.

$\alpha$ can then be written by
\begin{equation}
\alpha = \sum_{i,j} \alpha_{ij} p_{i} d_{f} p_{j} + A.     \label {4.10}
\end{equation}
As in Section 2, we introduce
\begin{equation}
a = \sum_{i,j} (1 + \alpha_{ij}) p_{i} d_{f} p_{j}.             \label{4.11}
\end{equation}
Notice that
\begin{equation}
a_{ii} = 1.                         \label{4.12}
\end{equation}
One then has
\begin{eqnarray}
a' &=& u a u^{*}, \nonumber   \\
a_{ij}^{'} &=& u(i) a_{ij} u(j)^{*}.                           \label{4.13}
\end{eqnarray}
 $A$ obeys the usual $U(1)$ gauge transformation rule,
\begin{equation}
  A' = A +u d u^{*}.                           \label{4.14}
\end{equation}
The curvature of the connection $\alpha$ is given by
\begin{equation}
\Theta = d \alpha  + \alpha^{2}.                             \label{4.15}
\end{equation}
It can be seen that $\Theta$ transforms 
in the usual way, $\Theta' = u \Theta u^{*}$.
As a $2$-form, $\Theta$ can be written as
\begin{eqnarray}
\Theta &=& d_{s} A + \sum_{ij} (d_{s} a_{ij}) p_{i} d_{f} p_{j}
 + \sum_{i,j,k} \theta_{ijk} p_{i} d_{f} p_{j} p_{j} d_{f} p_{k},  \nonumber \\
 \theta_{ijk} &=& a_{ij} a_{jk} - a_{ik}.              \label{4.16}
\end{eqnarray}
We see that $\Theta$ has a usual differential degree and a finit-difference
degree $(\alpha, \beta)$ adding up to $2$. Let us begin with the term 
in $\Theta$ of bi-degree $(2, 0)$:
\begin{equation}
\Theta^{(2, 0)} = F = d_{s} A.                               \label{4.17}
\end{equation}
$F$ is the field strength over the continous space-time $V$.

Next, we look at the component $\Theta^{(1, 1)}$ of bi-degree $(1, 1)$:
\begin{equation}
\Theta^{(1, 1)} = \sum_{ij} (d_{s} a_{ij}) p_{i} d_{f} p_{j}.  \label{4.18}
\end{equation}
$\Theta^{(1, 1)}$ corresponds to the interaction between $V$ and $M$.
It also obeys the field strength transformation rule:
\begin{equation}
\Theta'^{(1, 1)} = u \Theta^{(1, 1)} u^{*}.       \label{4.19}
\end{equation}

Notice that there is a peculiar property to the discrete space-time: 
 it is not $d_{s} + A$ but $d_{s}$ appears in (\ref{4.18})!
The reason is: we take $\alpha_{i} = A$ at the begining. Otherwise, there will
be a contribution from $\alpha_{i}$'s. This coincides with \cite{Co2} 
(pp.561-576) and \cite{Coq}
(where we replace $U(2)$ by $U(1)$ and take $\omega_{a} = \omega_{b}$ in 
\cite{Co2} and take $A =B$ in \cite{Coq}).

Finally, we have the component $\Theta^{(0, 2)}$ of degree $(0, 2)$:
\begin{equation}
\Theta^{(0, 2)} = \sum_{i,j,k} \theta_{ijk} 
                   p_{i} d_{f} p_{j} p_{j} d_{f} p_{k}.  \label{4.20}
\end{equation}
$\Theta^{(0, 2)}$ corresponds to the field strength over the finit space $M$. 

Just as in Section 3, we use the formula  (\ref{3.1}) to deal with the 
finite-difference degrees, i.e.,
\begin{equation}
 \pi(d_{f} p_{i}) = i[D, \pi(p_{i})].                     \label{4.21}
\end{equation}
We then obtain the action functinal over the discrete space-time $V \times M$:
\begin{equation} 
S = \int_{V} {\cal L}  d \nu.                         \label{4.22}
\end{equation}
The Lagrangian density is given by the following formulas:
\begin{equation} 
{\cal L} = {\cal L}_{2} + {\cal L}_{1} + {\cal L}_{0},   \label{4.23}
\end{equation}
\begin{equation} 
{\cal L}_{2} = \Vert\Theta^{(2, 0)}\Vert^{2} = \vert F \vert^{2} 
       = \vert d_{s} A \vert^{2},                               \label{4.24}
\end{equation}
\begin{equation}
{\cal L}_{1} = \Vert\Theta^{(1, 1)}\Vert^{2} = tr(d_{s} H)^{2}  
      = \frac{n}{n-1} \sum_{i} (d_{s}\phi)^{2}_{i},           \label{4.25}
\end{equation}
\begin{eqnarray}
{\cal L}_{0} &=& trH^{4} -2{\mu}^{2} trH^{2} +n{\mu}^{4} \nonumber \\
 &=& \sum_{i,j,k,l} (\sum_{\alpha} e^{\alpha}_{i} e^{\alpha}_{j} e^{\alpha}_{k}
     e^{\alpha}_{l}) \phi_{i} \phi_{j} \phi_{k} \phi_{l}
    -\frac{2n}{n-1} {\mu}^{2} (\sum_{i} \phi^{2}_{i}) + n {\mu}^{4}. 
                                                               \label{4.26}
\end{eqnarray}

{\bf 5. Geometric origin of phase transitions}

It is well know that the Landau-Ginzburg model is one of the most important
models in the theory of phase transitions. It is a kind of `metamodel' -
it captures the essence of many models in this field.

The Hamiltonian density of the Landau-Ginzburg model is
\begin{equation}
{\cal H}_{LG} =\frac{1}{2} \alpha^{2}\vert \nabla \phi\vert^{2}        
                 +\frac{1}{2}\mu^{2} \phi^{2}
                 +\frac{1}{4!}\lambda (\phi^{2})^{2},            \label{5.1}
\end{equation}
where $\phi$ is the order parameter, $\alpha,\mu^{2}, \lambda $  are 
phenomenological parameters. Notice that the temperature-dependent coefficient      
is $\mu^{2}$. This notation is not intended to imply that $\mu^{2} > 0$.

Consider now the case when $V \times M = V \times \{2-$point space$\}$, 
it is easy to see from equations (\ref{4.25}) and (\ref{4.26}) that 
${\cal L}_{1}$ and ${\cal L}_{0}$ correspond to the kinetic energy  and 
the potential term of ${\cal H}_{LG}$ below the critical temperature 
respectively. 

Moreover, in the continuous-spin formulation of the Ising model, see for example 
\cite{Bin,Iva}, the effective Hamiltonian can be expressed as ${\cal H}_{LG}$
plus additional terms, and these additional terms are of no consequence if 
all one seeks is critical exponents. Notice that the Ising model is just
the $2$-state Potts model. One then can imagine that the effective 
Hamiltonian of the continuous-spin formulation of the $n$-state Potts model 
\cite{Zia} may be related the action functional of $U(1)$ gauge field we 
derived in the previous section.

The effective Hamiltonian density of the $n$-state Potts model is
\begin{eqnarray}
{\cal H}_{P} &=& \sum_{i}(\nabla \phi_{i})^{2}
    +C_{1}(\sum_{i} \phi^{2}_{i})
 +C_{2}\sum_{i,j,k} (\sum_{\alpha} e^{\alpha}_{i} e^{\alpha}_{j} e^{\alpha}_{k})
             \phi_{i} \phi_{j} \phi_{k}  \nonumber\\
& &\mbox{}
+C_{3}\sum_{i,j,k,l} (\sum_{\alpha} e^{\alpha}_{i} e^{\alpha}_{j} e^{\alpha}_{k}
     e^{\alpha}_{l}) \phi_{i} \phi_{j} \phi_{k} \phi_{l}
    +C_{4}(\sum_{i} \phi^{2}_{i})^{2},                      \label{5.2}
\end{eqnarray}
where $C_{i} (i=1,2,3,4)$ is a constant. 

From equations (\ref{4.25}) and (\ref{4.26}), ${\cal L}_{1}$ really
corresponds to the kinetic energy term of ${\cal H}_{P}$, ${\cal L}_{0}$ 
is contained in the potential term of ${\cal H}_{P}$.  It might be asked:
why there is such a correspondance? This is because that the set of states 
of the Potts model forms a $n$ point space, one then can build the gauge 
theory on $\{$continuous space$\}$ $\times$ $\{n$-point space$\}$.  

In the Landau-Ginzburg model, the order parameter $\phi$ can be generalized
to the vector order paramter $\bf{\Phi}$. Obviously, the action functional 
of $U(1)$ gauge theory over $\{$continuous space$\}$ $\times$ 
$\{n$-point space$\}$ is a generalized Landau-Ginzburg Hamiltonian below 
the critical temperature. This means that there may exist a nontrivial
geometrical structure behind a class of phase transitions.

{\bf Acknowledgements}

The author would like to thank Profs. M.-L. Ge, M.-Z. Guo, Z.-J. Liu,
M. Qian and Z.-D. Wang for the helpful discussions.
This work was supported in part by the National Science Foundation 
of China and China Postdoctoral Science Foundation.

\end{document}